\documentclass[aps,pra,twocolumn,preprintnumbers,nofootinbib,amsmath,amssymb,footinbib]{revtex4}
\usepackage{mathtools}

\DeclareMathAlphabet{\mathcalligra}{T1}{calligra}{m}{n}
\DeclareMathAlphabet{\mathpzc}{OT1}{pzc}{m}{it}

\usepackage{graphicx,epsfig}
\usepackage{bm}
\usepackage{dcolumn}
\usepackage{xcolor}
\usepackage[breaklinks=true,colorlinks,citecolor=blue,linkcolor=blue,urlcolor=blue]{hyperref}

\def\nat#1#2#3{Nature {\bf #1}, #2 (#3)}

\def\sc#1#2#3{Science {\bf #1}, #2 (#3)}

\def\rmp#1#2#3{Rev. Mod. Phys. {\bf #1}, #2 (#3)}
\def\prl#1#2#3{Phys. Rev. Lett. {\bf #1}, #2 (#3)}
\def\pra#1#2#3{Phys. Rev. A {\bf #1}, #2 (#3)}

\def\pre#1#2#3{Phys. Rev. E {\bf #1}, #2 (#3)}

\def\epjb#1#2#3{Eur. Phys. J. B {\bf #1}, #2 (#3)}

\def\njp#1#2#3{New J. Phys. {\bf #1}, #2 (#3)}

\def\jpb#1#2#3{J. Phys. B: At. Mol. Opt. Phys. {\bf #1}, #2 (#3)}

\def\noi{\noindent}
\def\bc{\begin{center}}
\def\ec{\end{center}}
\newcommand{\bea}{\begin{equation}}
\newcommand{\eea}{\end{equation}\noi}
\newcommand{\ber}{\begin{eqnarray}}
\newcommand{\eer}{\end{eqnarray}\noi}
\begin{document}
\title{Efficiency fluctuations in quantum heat engines with non-quasi-static adiabatic strokes}
\author{Bikram Keshari Behera}\email{22PHPH18@uohyd.ac.in}
\author{Shyamal Biswas}\email{sbsp@uohyd.ac.in}
\affiliation{School of Physics, University of Hyderabad, C.R. Rao Road, Gachibowli, Hyderabad-500046, India}

\date{\today}

\begin{abstract}
We have presented an analytical and numerical study of efficiency fluctuations in quantum Carnot and Otto engines with pistons operating under non-quasi-static adiabatic expansion and compression processes at constant speed, along with quasi-static isothermal (for the Carnot engine) or isochoric processes (for the Otto engine). We have considered the two-point measurement scheme during the adiabatic strokes in this regard. By considering each of the pistons as a single quantum particle in a 1-D box with a non-quasi-statically movable boundary, we capture the stochastic nature of the work output resulting from the unitary evolution of the quantum mechanical state of the system during non-equilibrium processes. Furthermore, we have explicitly shown how the piston speed during quantum non-equilibrium strokes introduces additional quantum fluctuations that modify the average efficiency of both the engines.

\end{abstract}

\pacs{05.70.Ln Nonequilibrium and irreversible thermodynamics, 05.30.-d Quantum statistical mechanics, 07.20.Pe Heat engines; heat pumps; heat pipes, 05.40.-a Fluctuation phenomena, random processes, noise, and Brownian motion, 05.90.+m Other topics in statistical physics, thermodynamics, and nonlinear dynamical systems}

\maketitle

\section{Introduction}
The study of quantum thermodynamics with an emphasis on quantum heat engine is in the forefront over the last few decades \cite{Seifert2012,Kosloff2014,Vinjanampathy2016,Binder2018,Talkner2020,Landi2021,Pekola2021,Bhattacharjee2021,Cangemi2024,Campaioli2024,Campbell2026} both from experimental \cite{Rossnagel2016,Peterson2019,Lindenfels2019,Klatzow2019,Lisboa2022,Koch2023,Denzler2024,Barontini2025,Hou2025,Uusnakki2026,Xue2026} and theoretical \cite{Scovil1959,Alicki1979,Kieu2004,Quan2007,Quan2009,Kim2011,Horodecki2013,Xiao2015,Zheng2016,Millen2016,Kosloff2017,Watanabe2017,Ding2018,Denzler2020,Agarwalla2021,Purkayastha2022,Monnai2025,Svidzinsky2026} point of views. Quantum fluctuations play an important role in a quantum heat engine (as well as refrigerator) and dominate over the thermodynamic fluctuations if the engine operates at low temperatures and if its size is small enough. Consequently, anomalous heat flow has been used very recently to design a quantum Otto cycle with indefinite causal order \cite{Xue2026}. Such an engine generates heat and achieves refrigeration simultaneously \cite{Xue2026}. Recently, a single-atom ($^{40}$Ca$^{+}$) heat engine has also been made in the Paul-trap with dimensions $\sim 50$ nm and temperatures $\sim10$ mK \cite{Rossnagel2016}. Above all, the efficiency of a quantum engine significantly fluctuates about its quasi-static value \cite{Denzler2020}. It has been recently experimentally observed that in a small length-scale ($\sim$ 50 nm to $\mu$m) and low temperatures ($\sim5$ mK), the quantum Otto engine \cite{Hou2025,Klatzow2019}, on an average, significantly outperforms the corresponding classical heat engine by about 17\%.  Incidentally, the study of the probability distribution of the efficiency becomes the prime interest for a quantum heat engine.

Although the Otto engine is completely distinct from the Carnot engine, they share a similarity in their adiabatic expansion and compression strokes. Thus, the unitary evolution in both the quantum Carnot engine and the quantum Otto engine can be considered the same. Although the Carnot engine has never been realized as a practical technological device, its theoretical study remains relevant for understanding the maximum efficiency of engines with real-world applications. The Otto engine, on the other hand, has practical applications--such as powering petrol-driven vehicles \cite{Haywood2018}. However, we are interested in both engines at the mesoscopic length scale in order to capture the effect of quantum fluctuations on their efficiencies.

There has been significant theoretical development in the field regarding the analysis of the efficiencies (or average efficiency) of various types of quantum heat engines using different working substances. Examples include a 3-level maser in a Carnot engine \cite{Scovil1959}, a qubit (2-level atom/system\footnote{Here, a two-level system could also be ``a Bose-Einstein condensate confined to the bottom two energy levels of a trapping potential'' \cite{Kieu2004}.}) in an Otto engine \cite{Kieu2004}, a qubit in a Carnot engine \cite{Quan2007}, a single particle in a 1D box and a single-mode radiation field in a cavity under quasi-static processes \cite{Quan2009}, a single molecule in a Szilard engine \cite{Kim2011}, a few particles in a potential well undergoing relaxation in a Carnot cycle \cite{Xiao2015}, and a 1D harmonic oscillator with both quasi-statically/adiabatically \cite{Zheng2016} and non-adiabatically \cite{Ding2018,Denzler2020} changing frequencies in an Otto engine, \textit{etc}. Efficiency fluctuations for a quantum Otto engine with a time-dependent harmonic oscillator as a working substance have also been theoretically analyzed recently \cite{Denzler2020}.

In contrast, the working substance of classical heat engines is typically a gas inside a cylinder with a movable piston. The quantum Jarzynski equality has already been theoretically analyzed for a single particle in such a quantum piston \cite{Quan2012}. Surprisingly, quantum thermodynamics has not yet been studied for a single particle in a quantum piston undergoing cycles with non-quasi-static adiabatic strokes and quasi-static isothermal/isochoric strokes. In the present study, we aim to address this gap from a theoretical perspective for both the quantum Carnot engine and the quantum Otto engine. Our primary goal is to theoretically determine the efficiency fluctuations in both quantum heat engines, specifically for non-quasi-static adiabatic strokes.

Although there is no bound on efficiency fluctuations in a quantum engine \cite{Denzler2020}, a lower bound exists for the ratio of work fluctuations to input heat fluctuations in a finite-time quantum Otto engine \cite{Agarwalla2021}. This ratio, however, can be determined by the square of the engine's average efficiency \cite{Agarwalla2021}. Building on these foundational results, we aim to theoretically determine the efficiency fluctuations in quantum Carnot and Otto engines using a ``quantum piston" model, in which a single quantum particle is confined within a 1D infinite square well with a movable wall \cite{Quan2012}. The motion of the wall renders the system's Hamiltonian time-dependent, driving quantum mechanical transitions among energy eigenstates via unitary evolution during the adiabatic expansion and compression strokes. The transition probabilities under this unitary evolution can be obtained from the exact analytical solution to the time-dependent Schrödinger equation for a single particle in a uniformly moving piston \cite{Doescher1969}. A quantum Carnot engine can operate through such non-quasi-static adiabatic strokes with uniform expansion and compression, followed by quasi-static isothermal strokes during which it is thermally coupled to hot and cold reservoirs at absolute temperatures $T_h$ and $T_c$ ($T_h > T_c$), respectively. Conversely, a quantum Otto engine operates using the same non-quasi-static adiabatic strokes, followed by quasi-static isochoric strokes corresponding to heat addition and rejection processes, respectively. In this context, we theoretically analyze non-quasi-static driving protocols by tuning the speed of the quantum piston. Furthermore, we determine the work fluctuations within the conventional two-point projective measurement scheme \cite{Denzler2020, Denzler2024}. This approach yields a general expression for the efficiency distribution that incorporates both thermal fluctuations—arising from the Boltzmann occupation probabilities of the single-particle quantized energy eigenstates--and quantum fluctuations, which govern transition probabilities among the system's energy eigenstates. For an adiabatic stroke of finite duration, quantum fluctuations dominate thermal fluctuations when the piston size is small and the temperature is low. This results in multiple infinitesimally narrow peaks of significant height at discrete efficiency values (reflecting the equality between work done and energy differences) in the efficiency distributions of both the quantum Carnot and quantum Otto engines. Consequently, this enables us to achieve our objective: characterizing efficiency fluctuations in quantum heat engines operating with non-quasi-static adiabatic strokes.

The article is organized as follows. We analytically determine the probability of a particle making a transition from one of its energy eigenstates to another in the unitary process of a uniformly expanding or contracting quantum piston. In Section 2, we obtain results for the quantum Carnot engine. In Section 2A, we obtain the transition probabilities for the unitary evolutions in the adiabatic expansion and compression strokes. In Section 2B, we determine the average heat inputs and stochastic work outputs in a cycle of the quantum Carnot engine according to the two-point measurement scheme employed therein. In Section 2C, we numerically find the probability distribution function for arbitrary work output from a cycle of the quantum Carnot engine. In Section 2D, we numerically find the efficiency distribution function for the quantum Carnot engine. In Section 2E, we analyze the average efficiency of the quantum Carnot engine for its different piston speeds. In Section 3, we find the efficiency distribution function for the quantum Otto engine and analyze its average efficiency for different piston speeds in a similar manner. Finally, we conclude in Section 4.

\section{Results for the quantum Carnot engine with non-quasi-static adiabatic strokes}
Let us consider the piston of the quantum Carnot engine to be as a particle of mass $M$ in a 1-D box (i.e. an infinite square-well) along the $x$-axis, with a fixed wall at $x=0$ and a moving wall at $x=\lambda(t)$ at time $t$. The Hamiltonian for the particle can take the time-dependent form $\hat{H}(t)=-\frac{\hbar^2}{2M}\frac{\text{d}^2}{\text{d}x^2}+V_\lambda(x)$ \cite{Doescher1969,Quan2012}, where $\lambda=\lambda(t)$ is an external time-dependent control parameter for the potential energy\footnote{Here, $V_\lambda(x)=0$ for $0<x<\lambda(t)$, and it is $\infty$ beyond this domain.} $V_\lambda(x)$ for the motion of the particle, and varies according to the processes/strokes undergone in the engine. The length of the piston at any instant of time in such a process can be denoted as $l(t)$.

\begin{figure}
\includegraphics[width=8.5cm]{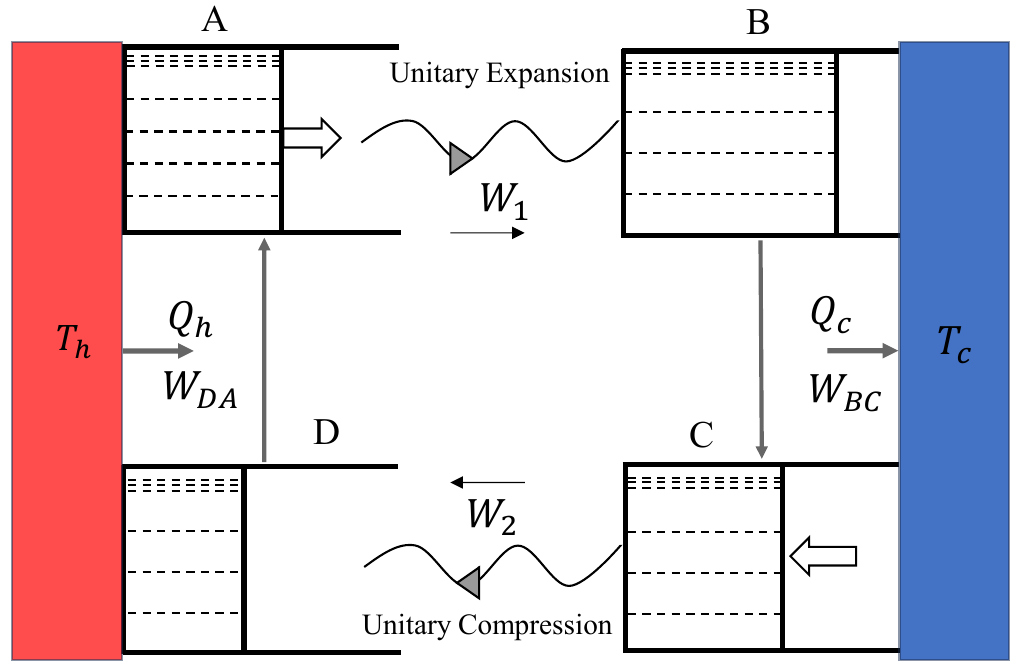}
\caption{Schematic of a four-stroke quantum Carnot engine, having isothermal expansion (D$\rightarrow$A) at temperature $T_h$, unitary expansion (A$\rightarrow$B), isothermal compression (B$\rightarrow$C) at temperature $T_c$, and unitary compression (C$\rightarrow$D).}
\label{fig2}
\end{figure}

The strokes of the quantum Carnot engine, in a cycle, as described in figure \ref{fig2}, are (i) hot quasi-static isothermal stroke (D$\rightarrow$A; with initial $l(t)=l_0'$ and final $l(t)=l_0>l_0'$) with an absorption of $Q_h$ amount of heat from the hot bath at a temperature $T_h$ and $W_{DA}>0$ amount of work done by the piston, (ii) non-quasi-static adiabatic (unitary) expansion stroke (A$\rightarrow$B; with an uniform speed $v$ for the time interval $\triangle t=\tau$) with an amount of work $W_1$ done by the system for an expansion from the initial length $l(t)=l_0$ to the final length $l(t)=l_\tau=l_0+v\tau$, (iii) cold quasi-static isothermal stroke (B$\rightarrow$C; with initial length $l(t)=l_\tau$ and final length $l(t)=l_\tau l_0'/l_0$) with a release of $Q_c$ amount of heat to the cold heat bath at a temperature $T_c<T_h$ and $W_{BC}<0$ amount of work done by the piston, and (iv) non-quasi-static adiabatic (unitary) compression stroke (C$\rightarrow$D; with an uniform speed $v$ for the time interval $\triangle t=\tau$) with an amount of work $W_2$ done by the system for an contraction from the initial length $l(t)=l_\tau l_0'/l_0$ to the final length $l(t)=l_0'$. 

Here, the final length ($l_t=l_\tau l_0'/l_0$) of the piston, in the cold quasi-static isothermal stroke (B$\rightarrow$C), is the classical correspondence ($V_A/V_D=V_C/V_B$\footnote{Here, $V_A$ means volume of the piston at the point $A$ and so on in the schematic diagram (figure {\ref{fig4}}).} \cite{Zemansky1997}) of the quantum adiabatic processes in the limiting case of the quasi-static expansion/contraction ($v\rightarrow0$, $\tau\rightarrow\infty$, and $v\tau=finite$) in the cycle of the Carnot engine. 

The two-point measurement scheme \cite{Talkner2007,Campisi2011} is applied only to the adiabatic expansion and compression strokes, in particular, during the initial and the final stage of the unitary evolution of the system (particle in the piston) in each of the adiabatic strokes. According to this scheme, the outcome of the measurement of the energy of the system in a full cycle can be described as follows. The system is found to be in thermodynamic equilibrium in contact with the hot heat bath for essentially all the time during the quasi-static (slow\footnote{Here, the thermalization time scale is faster than the time scale for the isothermal expansion.}) hot isothermal stroke (D$\rightarrow$A). The hot heat bath is then taken off from the system for a unitary evolution for a time interval $\triangle t=\tau$ in the non-quasi-static adiabatic expansion stroke (A$\rightarrow$B). The energy of the system is measured at the beginning and the final stages of this stroke, and the system is found to be in the energy eigenstates $|n_{l_0}\rangle$ with energy eigenvalue $E_n(l_0)=\frac{n^2\pi^2\hbar^2}{2Ml_0^2}$ (where $n\in\mathbb{N}$) and $|m_{l_\tau}\rangle$ with energy eigenvalue $E_m(l_\tau)=\frac{m^2\pi^2\hbar^2}{2Ml_\tau^2}$ (where $m\in\mathbb{N}$), respectively. Then the system is found to be in thermodynamic equilibrium in contact with the cold heat bath for essentially all the time during the quasi-static cold isothermal stroke (B$\rightarrow$C). The cold heat bath is then taken off from the system for a unitary evolution for a time interval $\triangle t=\tau$ in the non-quasi-static adiabatic compression stroke (C$\rightarrow$A). The energy of the system is measured at the beginning and the final stages of this stroke, and the system is found to be in the energy eigenstates $|j_{l_\tau l_0'/l_0}\rangle$ with energy eigenvalue $E_j(l_\tau l_0'/l_0)=\frac{j^2\pi^2\hbar^2}{2M[l_\tau l_0'/l_0]^2}$ (where $j\in\mathbb{N}$) and $|k_{l_0'}\rangle$ with energy eigenvalue $E_k(l_0')=\frac{k^2\pi^2\hbar^2}{2Ml_0'^2}$ (where $k\in\mathbb{N}$), respectively. The hot heat bath is then kept in contact with the system for a thermodynamic equilibrium in quasi-static manner to start the hot isothermal process (D$\rightarrow$A).

\subsection{Transition probabilities for the unitary evolutions in the adiabatic expansion and compression strokes}
Let us now analytically determine the transition probabilities for the unitary evolutions during the expansion and compression strokes as solutions to the time-dependent Schrödinger equation. Such a determination, of course, would be significantly different from that of a classical Carnot engine, from a quantum nonequilibrium dynamics point of view.

According to the adiabatic expansion stroke and the two-point measurement schemes, as described above,  the time-dependent Schrödinger equation results in the time evolution of the initial energy eigenstate $\phi_n(x,l_0)=\langle x|n_{l_0}\rangle=\sqrt{2/l_0}\sin(n\pi x/l_0)$, after the time internal $\triangle t=\tau$, as \cite{Doescher1969,Quan2012}
\begin{eqnarray}\label{eqn10}
\Phi_n(x,\tau)= \phi_n(x,l_{\tau})\text{e}^{i[\frac{1}{2}Mvx^2-E_n(l_0)l_0\tau]/\hbar l_{\tau}}.
\end{eqnarray}
The general solution to the time-dependent Schrödinger equation, of course, is $\psi(x,t)=\sum_{n=1}^{\infty}c_n\Phi_n(x,t)$ for $0\le t\le\tau$. It should be mentioned in this regard that the velocity-dependent extra phase terms in Eqn.~(\ref{eqn10}) represent the geometric phase due to the adiabatic motion of the piston. The time evolution of the initial energy eigenstate, however, undergoes with a unitary evolution operator $\hat{U}(0,\tau)$ such that $\Phi_n(x,\tau)=\langle x\vert n_{l_\tau}\rangle=\langle x\vert\hat{U}(0,\tau)\vert n_{l_0}\rangle$. The initial state $\vert n_{l_0}\rangle$ and the final state $\vert m_{l_\tau}\rangle$, under the two-point projective measurement scheme, are connected by the same unitary evolution operator such that $\hat{U}(0,\tau)$ such that $T_{n\rightarrow m}^{(0,\tau)}(l_0,l_\tau)=\vert \langle m_{l_\tau}\vert\hat{U}(0,\tau)\vert n_{l_0}\rangle\vert ^2$ represents the corresponding transition probability, as
\begin{eqnarray}\label{eqn12}
&&T_{n\rightarrow m}^{(0,\tau)}(l_0,l_\tau)=\bigg\vert \sum_{j=1}^{\infty}\frac{2}{l_0} \int_{0}^{l_0}\text{d}x\text{e}^{-i\frac{Mv x^2}{2\hbar l_0}}\sin \bigg(\frac{j \pi x}{l_0}\bigg)\times\nonumber\\&&\sin \bigg(\frac{n \pi x}{l_0}\bigg)\times\text{e}^{-i \frac{\hbar\pi^2 j^2 (l_{\tau}-l_0)}{2Ml_0 l_{\tau}v}}\nonumber\\&&\times\frac{2}{l_{\tau}} \int_{0}^{l_{\tau}} {\rm d}x'\text{e}^{i\frac{Mv x^2}{2\hbar l_{\tau}}} \sin \bigg( \frac{m \pi x'}{l_{\tau}}\bigg)\sin\left( \frac{j \pi x'}{l_{\tau}} \right) \bigg\vert^{2}.
\end{eqnarray}
To simplify our analysis, we assume that both the expansion and compression strokes are performed over the same time interval and that the piston velocity changes to a negative value for the compression stroke. 

The evaluation of the transition probability (Eqn.~(\ref{eqn12})) is not so difficult but tedious, as the integrals are similar and 1st one appears in terms of the oscillatory type functions, say Fresnel integrals, as $\sqrt{\frac{\pi\hbar l_0}{16Mv}}\big\{\text{e}^{i\frac{[j-n]^2\pi^2\hbar}{2l_0Mv}}\big[\{C\big(g_{j,n}(v)\big)-C\big(g_{j,n}(-v)\big)\}+i\{S\big(g_{j,n}(-v)\big)-S\big(g_{j,n}(v)\big)\}\big]+\text{e}^{i\frac{[j+n]^2\pi^2\hbar}{2l_0Mv}}\big[\{C\big(g_{j,-n}(-v)\big)-C\big(g_{j,-n}(v)\big)\}-i\{S\big(g_{j,-n}(-v)\big)-S\big(g_{j,-n}(v)\big)\}\big]\big\}$, where $g_{j,n}(v)=\frac{[j-n]\pi-Mvl_0/\hbar}{\sqrt{\pi M|v|l_0/\hbar}}$.

\begin{figure}
\includegraphics[width=8.5cm]{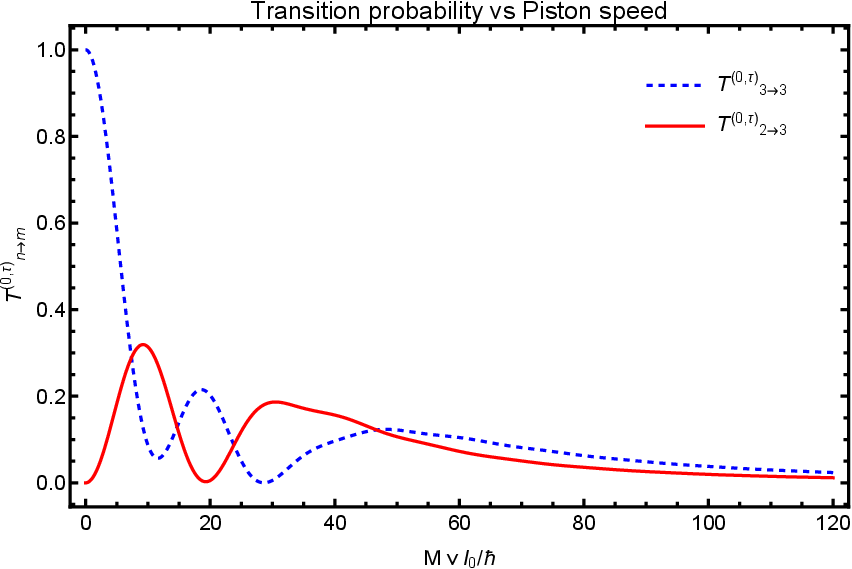}
\caption{Both the solid and dotted lines follow Eqn.~(\ref{eqn12}). The parameters for the solid line are set as $l_{\tau}/l_0=2,~\tau=0.2Ml_0^2/\hbar,~m=2,~n=3$, and $l_{\tau}/l_0=2,~\tau=0.2 Ml_0^2/\hbar,~m=3,~n=3$, respectively, with $l_0=~$unity.}
\label{fig3}
\end{figure}

With the above forms of the Fresnel integrals, we plot the transition probability (Eqn.~(\ref{eqn12})) with respect to the speed ($v$) for different pairs of the quantum numbers $n$ and $m$ in figure \ref{fig3}. It is clear from the figure that $T_{n\rightarrow m}^{(0,\tau)}=\delta_{n,m}$ as $v\rightarrow0$, as also obtained by \cite{Denzler2020, Agarwalla2021} for a time-dependent harmonic oscillator problem. The transition probability, of course, goes to $0$ as $v\rightarrow\infty$ for all possible values of $n$ and $m$ because of the strong fluctuations of the geometric phases ($-Mv x^2/2\hbar l_0$ and $Mv x^2/2\hbar l_\tau$) in all the Fresnel integrals in Eqn.~(\ref{eqn12}). The corresponding transition probability $T_{j\rightarrow k}^{(0,\tau)}(l'',l_0')$  for the adiabatic compression stroke can, however, be obtained from the same Eqn.~(\ref{eqn12}) with the replacements: $l_0\rightarrow l''$, $l_\tau\rightarrow l_0'$, $v\rightarrow -v$, $n\rightarrow j$, and $m\rightarrow k$. 

The above form of transition probability (Eqn.~(\ref{eqn12})), for both the adiabatic expansion and the adiabatic compression process, is necessary to determine the joint probability distribution for works and heats involved in a cycle of the quantum Carnot engine according to the two-point measurement scheme for the adiabatic expansion and compression strokes. This joint probability, in turn, results in the probability distribution for the efficiency of the quantum Carnot engine.  

\subsection{The heat inputs and work outputs in a cycle of the quantum Carnot engine according to the two-point measurement scheme}
The two-point measurement scheme \cite{Talkner2007,Campisi2011,Deffner2008} is widely used in the quantum domain to determine thermodynamic quantities, such as heat and work, as they correspond to compatible with quantum fluctuation relations \cite{Udo2012, Deffner2008}. Let us now determine the joint probability distribution for work and heat according to the strokes involved in the quantum Carnot engine and the two-point measurement scheme involved there in, as described above, for our system (particle in the piston). 

The work done by the system during the non-quasi-static adiabatic expansion and compression strokes, however, would be
\begin{eqnarray}\label{eqn9ac}
W_1=[E_n(l_0)-E_m(l_\tau)]~\text{and}~W_2=[E_j(l'')-E_k(l_0')],
\end{eqnarray}
respectively, according to the first law of thermodynamics and the projective measurement scheme mentioned above. Interesting, $W_1$ and $W_2$ become stochastic in nature due to the arbitrariness in the choice of the quantum numbers $\{n,m,j,k\}$ is a particular cycle of the quantum Carnot engine and they can take values beyond the classical bound $0\le\eta_{n,m,j,k}\le1$. This leads to a stochastic nature in the heat absorbed by the system ($Q_h$) during the hot isothermal stroke and the heat released by the system ($Q_c$) during the cold isothermal stroke, and the efficiency of the engine ($\eta_{n,m,j,k}$), due to the projective measurement scheme.

While the works done by the system during the (non-quasi-static) adiabatic strokes directly depend on the unitary evolutions and the corresponding transition probabilities $T_{n\rightarrow m}^{(0,\tau)}(l_0,l_\tau)$ and $T_{j\rightarrow k}^{(\tau,0)}(l'',l_0')$, they are stochastic in nature due to the arbitrariness in the choice of the quantum numbers $\{n,m,j,k\}$ is a particular cycle. The system, of course, is in a quantum non-equilibrium state during the non-static adiabatic expansion and compression strokes. The works done by the system during the (quasi-static) isothermal strokes, on the other hand, are thermally averaged out due to the influence of the heat baths. 

The average work done by the system ($\bar{W}_{DA}(T)=\int_{l_0'}^{l_0}\bar{F}_T(L)\text{d}L$) during the hot isothermal expansion stroke, however, can be determined, according to the averaging over force $\vec{F}_{n,T}(L)=\frac{\hbar^2}{2M}|\nabla\phi_n(n,x)|_{x=L}^2\hat{i}$ \cite{Biswas2010} acting on the moving wall (at $x=L$) in the state $\phi_n(x,L)=\sqrt{2/L}\sin(n \pi x/L)$ with the statistical (Boltzmann) probability\footnote{Here, $Z_{L}(T)=\sum_{n=1}^{\infty}\text{e}^{-E_n(L)/k_BT}=[\vartheta_3(0,\text{e}^{-\pi\lambda_{T}^2/4L^2})-1]/2$ is the partition function for the system (with the piston-length $L$) at temperature $T$, $\lambda_{T}=\sqrt{2\pi\hbar^2/Mk_BT}$ is the thermal de Broglie wavelength for the system at temperature $T$,  and $\vartheta_3(0,\pi^2\hbar^2\beta_h/2ML^2)$ is the Jacobi (elliptic) theta function of the 3rd kind  \cite{Dey2020}.} $P_n(L,T)=\text{e}^{-E_n(L)/k_BT}/Z_{L}(T)$ at the temperature $T=T_h$, and integrating it over $L$ from $L=l_0'$ to $L=l_0$, as
\begin{eqnarray}\label{eqn9a}
\bar{W}_{DA}(T_h)=\int_{l_0'}^{l_0}\frac{\pi^2\hbar^2}{ML^2}\frac{\text{e}^{-\pi\lambda_{T_h}^2/4L^2}\vartheta_3^{'}(0,\text{e}^{-\pi\lambda_{T_h}^2/4L^2})}{L[\vartheta_3(0,\text{e}^{-\pi\lambda_{T_h}^2/4L^2})-1]}\text{d}L~
\end{eqnarray}
where $\beta_h=1/k_BT_h$ and $\frac{\pi^2\hbar^2}{2ML^2}\text{e}^{-\pi\lambda_{T_h}^2/4L^2}\vartheta_3'(0,\pi\lambda_{T_h}^2/4L^2)$ is the derivative of the Jacobi elliptic theta function of the 3rd kind ($\vartheta_3(0,\pi\lambda_{T_h}^2/4L^2)$) with respect to $\beta_h$ \cite{Dey2020}. In the classical limit ($T_h\rightarrow\infty$), this amount of the (average) work done by the system takes the limiting value $W_{DA}(T_h)\rightarrow k_BT_h\ln(l_0/l_0')$ and it would be the same as the amount of the average heat absorbed by the system ($\bar{Q}_h$) at $T_h\rightarrow\infty$ because the average of energy of the becomes a function of only number of particles and the temperature for an ideal classical gas. This would not be the same for the quantum Carnot engine as the average energy of the system is also a function of piston-length now.  
In a similar way, the work by the system during the isothermal compression stroke would be
\begin{eqnarray}\label{eqn9b}
\bar{W}_{BC}(T_c)=-\int_{l_\tau}^{l''}\frac{\pi^2\hbar^2}{ML^2}\frac{\text{e}^{-\pi\lambda_{T_c}^2/4L^2}\vartheta_3^{'}(0,\text{e}^{-\pi\lambda_{T_c}^2/4L^2})}{L[\vartheta_3(0,\text{e}^{-\pi\lambda_{T_c}^2/4L^2})-1]}\text{d}L~
\end{eqnarray}
where $\beta_c=1/k_BT_c$ and $l''=l_\tau l_0'/l_0$. While at a very large temperature ($l''\gg\lambda_{T_c}$), the work by the system during the cold isothermal compression stroke takes the classical form $\bar{W}_{BC}(T_c)\simeq -k_BT_c\ln(l_0/l_0')$, at a very low temperature ($l''\ll\lambda_{T_c}$) it takes a non-classical form $W_{BC}(T_c)\simeq-\frac{\pi^2\hbar^2}{2Ml_\tau^2}[(l_0/l_0')^2-1]$ from Eqn.~(\ref{eqn9b}), as expected. This amount of the work done on the system, of course, would not be the same as the average amount of the heat released ($\bar{Q}_c$) to the cold bath. 

According to the projective measurement scheme, as described above, the heat absorbed by the system at $T=T_h$, however, can be obtained from the first law of thermodynamics, as
\begin{eqnarray}\label{eqn9aa}
Q_h=[E_n(l_0)-E_k(l_0')]+\bar{W}_{DA}(T_h).
\end{eqnarray}
Similarly, the heat released by the system at $T=T_c$ would be
\begin{eqnarray}\label{eqn9ab}
Q_c=-[E_j(l'')-E_m(l_{\tau})]-\bar{W}_{BC}(T_c).
\end{eqnarray}
It is interesting to note that both the heat absorbed by the system  ($Q_h$) and the heat released by the system  ($Q_c$) could be negative if at least we have $E_m(l_0)<E_m(l_{0}')$ and $E_k(l'')>E_j(l_{\tau})$ for a random choice of the quantum numbers $\{n,m,j,k\}$ leading to negative efficiency of the quantum Carnot engine. When both of them becomes -ve, the engine behaves like a refrigerator, and when either of them becomes -ve, the efficiency of the engine can become -ve. So, the efficiency of the quantum Carnot engine ($\eta_{n,m,j,k}$) is not necessarily restricted by $0\le\eta_{n,m,j,k}\le1$, rather it takes the range $-\infty<\eta_{n,m,j,k}<\infty$ according to the random choice of the quantum numbers $\{n,m,j,k\}$ within the projective measurement scheme, as indicated above in terms of works ($W_1, W_2$) done during the adiabatic expansion and compression strokes. 

\subsection{Joint probability distribution for an arbitrary work outputs from a cycle of the quantum Carnot engine}
Efficiency of the quantum Carnot engine $\eta_{n,m,j,k}$ under the projective measurement scheme as described above, however, depends on the average heat input\footnote{Here, the unitary evolution of the system corresponds to a mixed state, rather than a pure state, in contact with the heat bath. Thus, we need the average heat input even for the definition of the stochastic efficiency of the quantum Carnot engine.} ($\bar{Q}_h$) and the total work output $W=\bar{W}_{DA}(T_h)+W_1+\bar{W}_{BC}(T_c)+W_2$, according to Eqns.~(\ref{eqn9ac}), (\ref{eqn9a}), (\ref{eqn9b}), and (\ref{eqn9aa}). Here, the work output ($\bar{W}_{DA}(T_h)$ and $\bar{W}_{BC}(T_c)$) during the isothermal strokes, being averaged out, as mentioned above, are not stochastic in nature and, of course, are independent of the quantum numbers $\{n,m,j,k\}$. These two quantities can be treated as constants in the cycles of the quantum Carnot engine. Here $W_1$ and $W_2$ (so as $Q_h$ and $Q_c$ according to Eqns.~(\ref{eqn9aa}) and (\ref{eqn9ab}), respectively), are, of course, stochastic in nature as they depend on the arbitrary quantum numbers $\{n,m,j,k\}$ according to the projective measurement scheme. Thus, the joint probability of the stochastic input heat ($Q_h$), stochastic output heat ($Q_c$), and stochastic output works ($W_1$ and $W_2$) in a chain of the consecutive strokes in a particular cycle with the quantum numbers $\{n,m,j,k\}$, according to Eqns.~(\ref{eqn9ac}), (\ref{eqn9a}), (\ref{eqn9b}), (\ref{eqn9aa}) and (\ref{eqn12}), would be
\begin{eqnarray}\label{eqn9x}
&&P_{n,m,j,k}(Q_h,W_1,Q_c,W_2)=\delta\big(Q_h-[E_n(l_0)-E_k(l_0')]\nonumber\\&&-\bar{W}_{DA}(T_h)\big)
\times\delta\big(W_1-[E_n(l_0)-E_m(l_\tau)]\big)T_{n\rightarrow m}^{(0,\tau)}(l_0,l_\tau)\nonumber\\&&\times
\delta\big(Q_c+[E_j(l'')-E_m(l_{\tau})]+\bar{W}_{BC}(T_c)\big)\nonumber\\&&\times\delta\big(W_2-[E_j(l'')-E_k(l_0')]\big)T_{j\rightarrow k}^{(\tau,0)}(l'',l_0').
\end{eqnarray}
The heat absorbed by the system ($Q_h$) and heat released by the system ($Q_c$) during the hot and cold quasi-static isothermal strokes (respectively) are, however, averaged out, as the system remains in thermodynamic equilibrium with the hot/cold heat bath. Hence, the heat absorbed by the system and released by the system essentially cannot be random variables if we consider the isothermal strokes to be quasi-static. Thus, we simplify Eqn.~(\ref{eqn9x}) for the quasi-static isothermal strokes, with the replacement of $Q_h$ with $\bar{Q}_h$ and $Q_c$ with $\bar{Q}_c$, as
\begin{eqnarray}\label{eqn9}
&&P_{n,m,j,k}(W_1,W_2)=\delta\big(W_1-[E_n(l_0)-E_m(l_\tau)]\big)\times\nonumber\\&& T_{n\rightarrow m}^{(0,\tau)}(l_0,l_\tau)
\delta\big(W_2-[E_j(l'')-E_k(l_0')]\big)T_{j\rightarrow k}^{(\tau,0)}(l'',l_0'),~~~~
\end{eqnarray}
with the respective average values of the heat input and heat output, as
\begin{eqnarray}\label{eqn9y}
\bar{Q}_h&=&[\sum_{n=1}^{\infty}P_n(l_0,T_h)E_n(l_0)-\sum_{n=1}^{\infty}P_n(l_0',T_h)E_n(l_0')]\nonumber\\&&+\bar{W}_{DA}(T_h)\nonumber\\&=&\frac{\pi^2\hbar^2}{4Ml_0^2}\text{e}^{-\pi\lambda_{T_h}^2/4l_0^2}\vartheta_3^{'}(0,\text{e}^{-\pi\lambda_{T_h}^2/4l_0^2})+\bar{W}_{DA}(T_h)\nonumber\\&&-\frac{\pi^2\hbar^2}{4Ml_0'^2}\text{e}^{-\pi\lambda_{T_h}^2/4l_0'^2}\vartheta_3^{'}(0,\text{e}^{-\pi\lambda_{T_h}^2/4l_0'^2})
\end{eqnarray}
and
\begin{eqnarray}\label{eqn9z}
\bar{Q}_c&=&-[\sum_{n=1}^{\infty}P_n(l'',T_c)E_n(l'')-\sum_{n=1}^{\infty}P_n(l_\tau,T_c)E_n(l_\tau)]\nonumber\\&&-\bar{W}_{BC}(T_c)\nonumber\\&=&-\frac{\pi^2\hbar^2}{4Ml''^2}\text{e}^{-\pi\lambda_{T_c}^2/4l''^2}\vartheta_3^{'}(0,\text{e}^{-\pi\lambda_{T_c}^2/4l''^2})-\bar{W}_{BC}(T_c)\nonumber\\&&+\frac{\pi^2\hbar^2}{4Ml_\tau^2}\text{e}^{-\pi\lambda_{T_c}^2/4l_\tau^2}\vartheta_3^{'}(0,\text{e}^{-\pi\lambda_{T_c}^2/4l_\tau^2})
\end{eqnarray}
where $P_n(L,T)=\text{e}^{-E_n(L)/k_BT}/Z_{L}(T)$, as mentioned before, are the occupation (Boltzmann) probabilities of the states $\{|n_{L}\rangle\}$ of the system at temperatures $T=T_h$ and $T=T_c$, respectively, for various lengths of the piston $L=l_0',l_0,l_\tau,~\&~l''$.

If the cycle changes, then the quantum numbers $\{n,m,j,k\}$ may change randomly. Thus, the probability distribution for the given stochastic output works $W_1$ and $W_2$, after averaging over all possible cycles (i.e. over all possible quantum numbers $\{n,m,j,k\}$), can be obtained from Eqn.~(\ref{eqn9}), with the appropriate statistical weight of the initials energy eigenstates of the system during the adiabatic expansion and compression strokes, as
\begin{eqnarray}\label{eqn9d}
P(W_1,W_2)&=&\sum_{\{n,m,j,k\}}P_n(l_0,T_h)P_m(l_\tau,T_c)P_j(l'',T_c)\nonumber\\&&\times P_k(l_0',T_h) P_{n,m,j,k}(W_1,W_2).
\end{eqnarray}
We need all four statistical weight factors (i.e. the Boltzmann probabilities) in Eqn.~(\ref{eqn9}) because all the quantum numbers $\{n,m,j,k\}$ in the chain of the strokes also correspond to the energy eigenstates when the system is in thermodynamic equilibrium, especially, during the isothermal strokes.  The probability distribution function $P(W_1,W_2)$ (Eqn.~(\ref{eqn9d})) for the given stochastic variables $\{W_1,W_2\}$ is useful to determine the probability distribution for the efficiency of the quantum Carnot engine. Discrete spikes in the probability distribution (otherwise zero), of course, would appear for certain values of $\{W_1,W_2\}$ satisfying the conditions (\ref{eqn9ac}) for certain values of the discrete quantum numbers $\{n,m,j,k\}$. 

\subsection{Efficiency distribution function for the quantum Carnot engine}
As indicated above, the efficiency of the quantum Carnot engine becomes stochastic, as both the work output $W=\bar{W}_{DA}(T_h)+W_1+\bar{W}_{BC}(T_c)+W_2$ is stochastic in nature according to the the two-point measurement scheme, unlike the average values of heat input ($\bar{Q}_h$) and work output ($\bar{W}$) in the classical thermodynamics. The stochastic efficiency for every single realisation ($\{n,m,j,k\}$) in a cycle of the quantum Carnot engine can be defined, as explained above, as
\begin{eqnarray}\label{eqn14}
\eta_{n,m,j,k}=\frac{\bar{W}_{DA}(T_h)+W_1+\bar{W}_{BC}(T_c)+W_2}{\bar{Q}_h},
\end{eqnarray}
where $\bar{W}_{DA}(T_h)$, $\bar{W}_{BC}(T_c)$, and $\bar{Q}_h$ follow Eqns.~(\ref{eqn9a}), (\ref{eqn9b}), and (\ref{eqn9y}), respectively. Here, by definition, $\eta_{n,m,j,k}\in\mathbb{R}$ is a stochastic variable due to the presence of the stochastic variables $\{W_1,W_2\}$ in Eqn.~(\ref{eqn14}), and it, of course, has the range $-\infty<\eta_{n,m,j,k}<\infty$ for all possible values of the quantum numbers $\{n,m,j,k\}$. 

According to the values of the quantum numbers $\{n,m,j,k\}$, each of the variables $\{W_1,W_2\}$ can be safely taken as a continuous variable in $\mathbb{R}$ by virtue of the probability distribution function $P(W_1,W_2)$ (Eqn.~(\ref{eqn9d})). The efficiency can, however, be made a continuous variable ($\eta$) irrespective of its value in a particular cycle, by virtue of its probability distribution function $P(\eta)$, as
\begin{eqnarray}\label{eqn15}
P(\eta)&=&\int_{-\infty}^{\infty}\int_{-\infty}^{\infty}\text{d}W_1\text{d}W_2P(W_1,W_2)\times\nonumber\\&&\delta\bigg(\eta-\frac{\bar{W}_{DA}(T_h)+W_1+\bar{W}_{BC}(T_c)+W_2}{\bar{Q}_h}\bigg),~~~~~
\end{eqnarray}
where the joint probability distribution $P(W_1,W_2)$, as obtained in Eqn.~(\ref{eqn9d}), takes care of all the possible outcomes of the two-point measurements and their respective probabilities. This probability distribution function for the (stochastic) efficiency, of course, is the normalized efficiency distribution function. Evaluation of the integrations in Eqn.~(\ref{eqn15}), according to the properties of the Dirac deltas, recast the normalized efficiency distribution function for the quantum Carnot engine, as 
\begin{eqnarray}\label{eqn16}
&&P(\eta)=\sum_{\{n,m,j,k\}}\frac{\text{e}^{-\frac{E_n(l_0)}{k_BT_h}}}{Z_{l_0}(T_h)}\frac{\text{e}^{-\frac{E_m(l_\tau)}{k_BT_c}}}{Z_{l_\tau}(T_c)}\frac{\text{e}^{-\frac{E_j(l'')}{k_BT_c}}}{Z_{l''}(T_c)}\frac{\text{e}^{-\frac{E_k(l_0')}{k_BT_h}}}{Z_{l_0'}(T_h)}\nonumber\\&&\times T_{n\rightarrow m}^{(0,\tau)}(l_0,l_\tau)T_{j\rightarrow k}^{(\tau,0)}(l'',l_0')\delta\bigg(\eta-\bigg[\bar{W}_{DA}(T_h)+[E_n(l_0)\nonumber\\&&-E_k(l_0')]+\bar{W}_{BC}(T_c)+[E_j(l'')-E_m(l_\tau)]\bigg]/\bar{Q}_h\bigg).\nonumber\\
\end{eqnarray}
Is is interesting to note that the efficiency distribution function, as derived in Eqn.~(\ref{eqn16}), depends on both the size of the piston\footnote{It becomes independent of the size only in the classical limit.} and the temperatures $T_c$ and $T_h$. The effect of the quantum and thermal fluctuations on the efficiency distribution function are captured by all the four Boltzmann probabilities and the corresponding quantized energy eigenvalues in Eqn.~(\ref{eqn15}).  Additional effect of the quantum fluctuations on the efficiency distribution functions due to the motion of the wall are also captured by the quantum mechanical transition probabilities in the same equation.

\subsection{Average efficiency of the quantum Carnot engine}
The average of the efficiency ($\bar{\eta}=\int_{-\infty}^{\infty}\eta P(\eta)\text{d}\eta$) can, however, be obtained from Eqn.~(\ref{eqn16}), as
\begin{eqnarray}\label{eqn16a}
\bar{\eta}&=&\sum_{\{n,m,j,k\}}\frac{\text{e}^{-\frac{E_n(l_0)}{k_BT_h}}}{Z_{l_0}(T_h)}\frac{\text{e}^{-\frac{E_m(l_\tau)}{k_BT_c}}}{Z_{l_\tau}(T_c)}\frac{\text{e}^{-\frac{E_j(l'')}{k_BT_c}}}{Z_{l''}(T_c)}\frac{\text{e}^{-\frac{E_k(l_0')}{k_BT_h}}}{Z_{l_0'}(T_h)}\nonumber\\&&\times T_{n\rightarrow m}^{(0,\tau)}(l_0,l_\tau)T_{j\rightarrow k}^{(\tau,0)}(l'',l_0')\times\nonumber\\&&\bigg[\bar{W}_{DA}(T_h)+[E_n(l_0)-E_k(l_0')]+\bar{W}_{BC}(T_c)\nonumber\\&&+[E_j(l'')-E_m(l_\tau)]\bigg]/\bar{Q}_h.
\end{eqnarray}
In the quasi-static limit ($v\rightarrow0$, $\tau\rightarrow\infty$, $v\tau=finite$), the transition probabilities in Eqn.~(\ref{eqn16a}) take the form $T_{n\rightarrow m}^{(0,\tau)}(l_0,l_\tau)=\delta_{n,m}$ and $T_{j\rightarrow k}^{(\tau,0)}(l'',l_0')=\delta_{j,k}$. Thus, we recast Eqn.~(\ref{eqn16a}) for the quasi-static case, as
\begin{eqnarray}\label{eqn16b}
\bar{\eta}_{qs}&=&\sum_{\{n,j\}}\frac{\text{e}^{-\frac{E_n(l_\tau)}{k_BT_h}}}{Z_{l_\tau}(T_h)}\frac{\text{e}^{-\frac{E_j(l'')}{k_BT_c}}}{Z_{l''}(T_c)}\bigg[1\nonumber\\&&-\frac{[E_n(l_\tau)-E_j(l'')]-\bar{W}_{BC}(T_c)}{\bar{Q}_h}\bigg]\nonumber\\&=&1-\frac{\bar{Q}_c}{\bar{Q}_h},
\end{eqnarray}
where $\bar{Q}_c$ follows Eqn.~(\ref{eqn9z}). It is interesting to note that the average efficiency  (Eqn.~(\ref{eqn16b})) of the quantum Carnot engine in the quasi-static limit, takes the the same form for the classical Carnot engine. This efficiency, according to Eqns.~(\ref{eqn9y}) and (\ref{eqn9z}), however, takes the well-known form $\bar{\eta}_{qs}\simeq1-T_c/T_h$ only in the classical regime ($\lambda_{T_c}\ll l_0'$). 

\begin{figure}
\includegraphics[width=8.5cm]{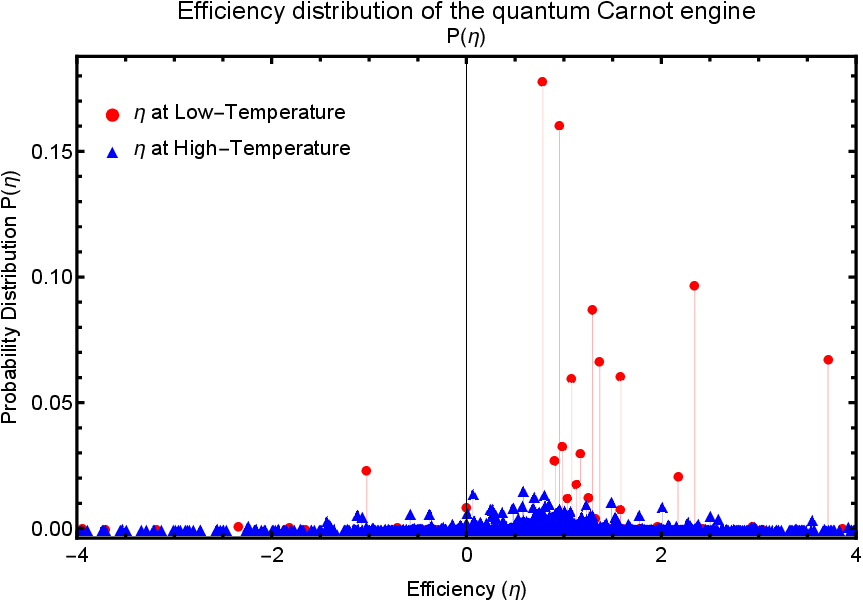}
\caption{The plots represent efficiency distribution for the quantum Carnot engine and follow Eqn.~(\ref{eqn16}) for $v=5 \hbar/Ml_0$, $\tau=0.2 M l_0^2/\hbar$, and $l_0=unity$ (i) at low temperatures with $\beta_c=8 (2 M l_0''^2/\pi^2 \hbar^2)$ and $\beta_h=0.3 (2 M l_0^2/\pi^2 \hbar^2)$ (circles), and (ii) at high temperatures with $\beta_c=0.02 (2 M l_0''^2/\pi^2 \hbar^2)$ and $\beta_h=0.01(2 M l_0^2/\pi^2 \hbar^2)$ (triangles).
}
\label{fig6}
\end{figure}

We plot the efficiency distribution function (Eqn.(\ref{eqn16})) for various temperature regimes in figure \ref{fig6}. The spikes in the figure arise from the matching conditions (Eqn.~(\ref{eqn14})) for various sets of quantum numbers. It is interesting to note that the efficiency fluctuations of the quantum Carnot engine are washed out in the high-temperature regime. In this regime, the quantum Carnot engine behaves like a classical Carnot engine.

\section{Results for the quantum Otto engine with non-quasi-static adiabatic strokes}
\begin{figure}
\centering
\includegraphics[width=8.5cm]{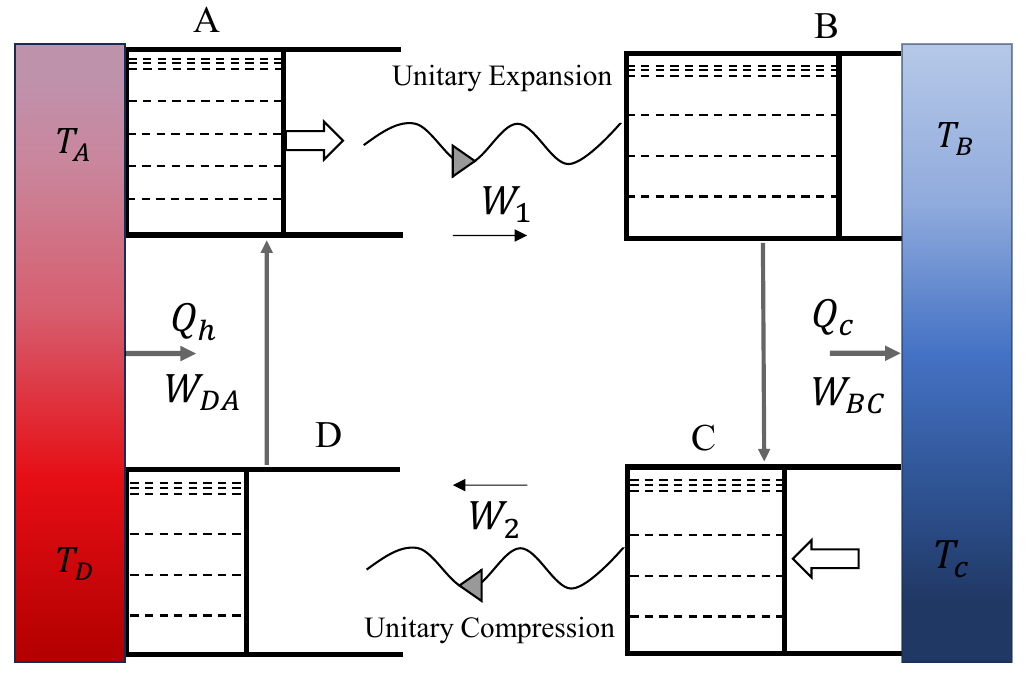}
\caption{Schematic of a four-stroke quantum Otto engine, having isochoric quasi-static combustion stroke with increase of temperature from  $T_D$ to $T_A$ ($D\rightarrow A$), non-quasi-static-unitary expansion with finite piston speed ($A\rightarrow B$), isochoric quasi-static heat rejection with temperature drop from $T_B$ to $T_C$ ($B\rightarrow C$), and non-quasi-static-unitary compression stroke with finite piston speed ($C\rightarrow D$).}
\label{fig1}
\end{figure}

The intake and exhaust strokes corresponding to isobaric expansion and contraction processes, respectively, don't directly take part in determining the efficiency of the Otto engine because the work done (and corresponding heat absorbed) by the system in two strokes cancel each other. Its compression stroke, on the other hand, accompanies two processes: adiabatic compression and isochoric combustion, which take part in determining the efficiency. Its power stroke consists of two processes: adiabatic expansion and isochoric heat rejection, which also take part in determining the efficiency. So, the relevant processes in a cycle, which determine the efficiency of the quantum Otto engine, as described in figure \ref{fig1}, are (i) isochoric quasi-static combustion process (D$\rightarrow$A; with $l(t)=l_0$) with an absorption of $\bar{Q}_h$ amount of heat and an increase in temperature from $T_D$ to $T_A$, (ii) non-quasi-static adiabatic (unitary) expansion process (A$\rightarrow$B; with an uniform speed $v$ for the time interval $\triangle t=\tau$) with an amount of work $W_1$ done by the system for an expansion from the initial length $l(t)=l_0$ to the final length $l(t)=l_\tau=l_0+v\tau$, (iii) isochoric quasi-static heat rejection process (B$\rightarrow$C; with length $l(t)=l_\tau$) with a release of $Q_c$ amount of heat to the surroundings and a drop in temperature from $T_B$ to $T_C$, and (iv) non-quasi-static adiabatic (unitary) compression stroke (C$\rightarrow$D; with an uniform speed $v$ for the time interval $\triangle t=\tau$) with an amount of work $W_2$ done by the system for an contraction from the initial length $l(t)=l_\tau$ to the final length $l(t)=l_0$. Hamiltonian of the system $\hat{H}(t)=-\frac{\hbar^2}{2M}\frac{\text{d}^2}{\text{d}x^2}+V_\lambda(x)$ remains the same as that for the quantum Carnot engine. The two-point projective measurement scheme within the adiabatic processes in the quantum Otto engine also remains the same as that in the quantum Carnot engine with a mapping $l_0'=l_0$ and $l''=l_\tau$.

\subsection{Efficiency distribution function for the quantum Otto engine}
According the processes involved in a cycle of the Otto engine, as described above, the average heat absorbed during the isochoric quasi-static (equilibrium) combustion process, according to the first law of thermodynamics, is
\begin{eqnarray}\label{eqn18}
\bar{Q}_h&=&[\sum_{n=1}^{\infty}P_n(l_0,T_A)E_n(l_0)-\sum_{n=1}^{\infty}P_n(l_0,T_B)E_n(l_0)]\nonumber\\&=&\frac{\pi^2\hbar^2}{4Ml_0^2}\text{e}^{-\pi\lambda_{T_A}^2/4l_0^2}\vartheta_3^{'}(0,\text{e}^{-\pi\lambda_{T_A}^2/4l_0^2})\nonumber\\&&-\frac{\pi^2\hbar^2}{4Ml_0^2}\text{e}^{-\pi\lambda_{T_D}^2/4l_0^2}\vartheta_3^{'}(0,\text{e}^{-\pi\lambda_{T_D}^2/4l_0'^2}).
\end{eqnarray}
Similarly, the heat rejected by the system during the isochoric quasi-static heat rejection process, would be
\begin{eqnarray}\label{eqn19}
\bar{Q}_c&=&\frac{\pi^2\hbar^2}{4Ml_0^2}\bigg[\text{e}^{-\pi\lambda_{T_B}^2/4l_0^2}\vartheta_3^{'}(0,\text{e}^{-\pi\lambda_{T_B}^2/4l_0^2})\nonumber\\&&-\text{e}^{-\pi\lambda_{T_C}^2/4l_0^2}\vartheta_3^{'}(0,\text{e}^{-\pi\lambda_{T_C}^2/4l_0'^2})\bigg].
\end{eqnarray}

\begin{figure}
\includegraphics[width=8.5cm]{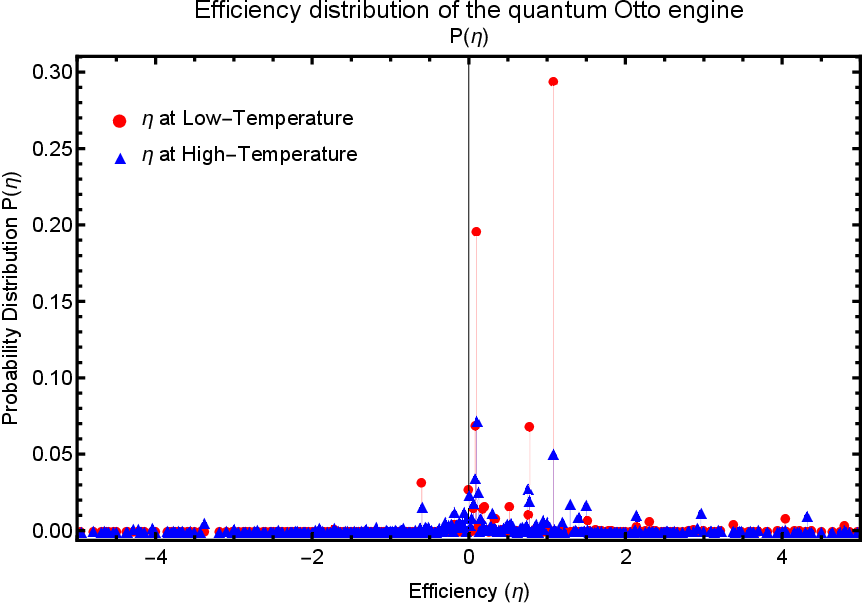}
\caption{The plots represent efficiency distribution function for the quantum Otto engine and follow Eqn.~(\ref{eqn17}) for $v=5 \hbar/Ml_0$, $\tau=0.2 M l_0^2/\hbar$, and $l_0=unity$ (i) at low temperatures with $\beta_A=8(2Ml_\tau^2/\pi^2 \hbar^2)$, $\beta_B=32(2Ml_0^2/\pi^2 \hbar^2)$, $\beta_C=0.3(2Ml_0^2/\pi^2 \hbar^2)$, and $\beta_D=0.08(2Ml_0^2/\pi^2 \hbar^2)$ (circles), and at high temperatures with $\beta_A=0.02(2Ml_\tau^2/\pi^2 \hbar^2)$, $\beta_B=0.18(2Ml_0^2/\pi^2 \hbar^2)$, $\beta_C=0.01(2Ml_0^2/\pi^2 \hbar^2)$, and $\beta_D=0.04(2Ml_0^2/\pi^2 \hbar^2)$ (triangles).}
\label{fig4}
\end{figure}

The (stochastic) work done by the system during the non-quasi-static adiabatic expansion and compression strokes ($W_1$ and $W_2$, respectively) in the quantum Otto engine, are of course individually the same as that for the quantum Carnot engine, and have already been mentioned in Eqn.~(\ref{eqn9ac}). Thus, we get the (stochastic) efficiency distribution function for the quantum Otto engine, in analogy with Eqn.~(\ref{eqn16}), by setting $\bar{W}_{DA}(T_h)=\bar{W}_{BC}(T_c)=0$ under the consideration of Eqns.~(\ref{eqn18}), (\ref{eqn19}), (\ref{eqn9ac}),  as
\begin{eqnarray}\label{eqn17}
&&P(\eta)=\sum_{\{n,m,j,k\}}\frac{\text{e}^{-\frac{E_n(l_0)}{k_BT_A}}}{Z_{l_0}(T_A)}\frac{\text{e}^{-\frac{E_m(l_\tau)}{k_BT_B}}}{Z_{l_\tau}(T_B)}\frac{\text{e}^{-\frac{E_j(l_\tau)}{k_BT_C}}}{Z_{l_\tau}(T_C)}\frac{\text{e}^{-\frac{E_k(l_0)}{k_BT_D}}}{Z_{l_0}(T_D)}\nonumber\\&&\times T_{n\rightarrow m}^{(0,\tau)}(l_0,l_\tau)T_{j\rightarrow k}^{(\tau,0)}(l_\tau,l_0)\delta\bigg(\eta-\bigg[[E_n(l_0)-E_k(l_0)]\nonumber\\&&+[E_j(l_\tau)-E_m(l_\tau)]\bigg]/\bar{Q}_h\bigg).
\end{eqnarray}
The classical correspondence of the above results further requires ($T_A/T_D=T_B/T_C$) \footnote{See the Sec. 6.2 of \cite{Zemansky1997}.}.

\subsection{Average efficiency of the quantum Otto engine}
The average of the efficiency ($\bar{\eta}=\int_{-\infty}^{\infty}\eta P(\eta)\text{d}\eta$) can now be obtained from Eqn.~(\ref{eqn17}), as
\begin{eqnarray}\label{eqn20}
\bar{\eta}&=&\sum_{\{n,m,j,k\}}\frac{\text{e}^{-\frac{E_n(l_0)}{k_BT_A}}}{Z_{l_0}(T_A)}\frac{\text{e}^{-\frac{E_m(l_\tau)}{k_BT_B}}}{Z_{l_\tau}(T_B)}\frac{\text{e}^{-\frac{E_j(l_\tau)}{k_BT_C}}}{Z_{l_\tau}(T_C)}\frac{\text{e}^{-\frac{E_k(l_0)}{k_BT_D}}}{Z_{l_0}(T_D)}\nonumber\\&&\times T_{n\rightarrow m}^{(0,\tau)}(l_0,l_\tau)T_{j\rightarrow k}^{(\tau,0)}(l_\tau,l_0)\times\nonumber\\&&\frac{[E_n(l_0)-E_k(l_0)]+[E_j(l_\tau)-E_m(l_\tau)]}{\bar{Q}_h}.
\end{eqnarray}
In the quasi-static limit ($v\rightarrow0$, $\tau\rightarrow\infty$, and $v\tau=finite$), Eqn.~(\ref{eqn17}) takes the form
\begin{eqnarray}\label{eqn21}
\bar{\eta}_{qs}&=&\sum_{\{n,j\}}\frac{\text{e}^{-\frac{E_n(l_\tau)}{k_BT_B}}}{Z_{l_\tau}(T_B)}\frac{\text{e}^{-\frac{E_j(l\tau)}{k_BT_C}}}{Z_{l_\tau}(T_C)}\bigg[1-\frac{[E_n(l_\tau)-E_j(l_\tau)]}{\bar{Q}_h}\bigg]\nonumber\\&=&1-\frac{\bar{Q}_c}{\bar{Q}_h},
\end{eqnarray}
where $\bar{Q}_c$ follows Eqn.~(\ref{eqn19}). It is interesting to note that the average efficiency  (Eqn.~(\ref{eqn16b})) of the quantum Otto engine in the quasi-static limit, also takes the same form as that for the classical Carnot engine. This efficiency, according to Eqns.~(\ref{eqn19}) and (\ref{eqn18}), however, takes the well-known form $\bar{\eta}_{qs}\simeq1-[T_B-T_C]/[T_A-T_D]$ only in the classical regime ($\lambda_{T_D}\ll l_0$). 

\begin{figure}
	\includegraphics[width=8.4cm]{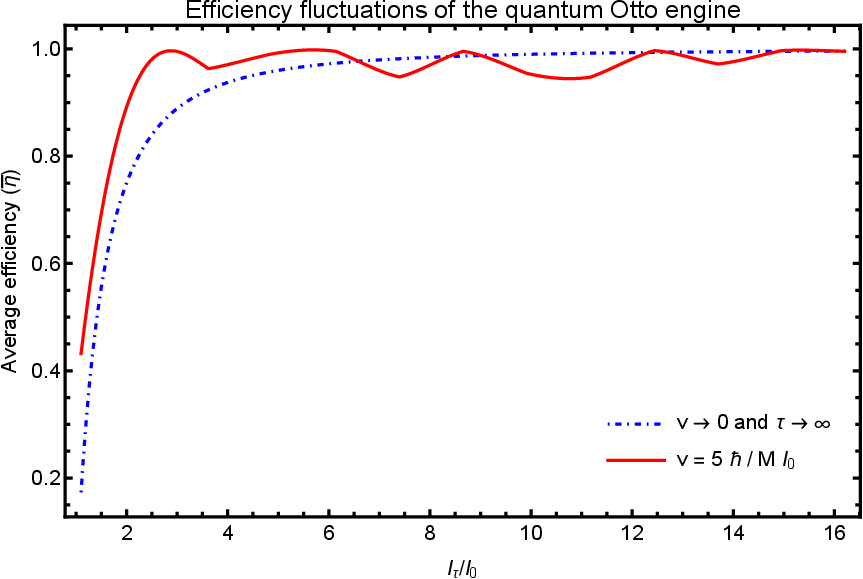}
	\caption{Plots follow Eqn.~(\ref{eqn20}) for the parameters $v=5\hbar/Ml_0$, $\tau=0.2Ml_0^2/\hbar$, and $l_0=unity$ (solid line); and $v\rightarrow0$, $\tau\rightarrow\infty$, and $v\tau=l_0=unity$ (dotted line).}
	\label{fig5}
\end{figure}

We plot the efficiency distribution function (Eqn.(\ref{eqn17})) for various temperature regimes in figure \ref{fig4}. The spikes in the figure arise due to the matching conditions (\textit{Eqn.~(\ref{eqn14}) with $\bar{W}_{DA}=\bar{W}_{BC}$ for $\bar{Q}_h$ of Eqn.~(\ref{eqn9y}) of Eqn.~(\ref{eqn9y})}) for various sets of quantum numbers $\{n,m,j,k\}$. Furthermore, as expected, the efficiency fluctuations of the quantum Carnot engine vanish in the high-temperature regime. In this regime, the quantum Otto engine behaves like a classical Carnot engine. We show the piston-length dependence of the quantum Otto engine (Eqn.(\ref{eqn20})) in figure \ref{fig5} for various speeds. The oscillations in the mean efficiency stem from the additional effect of quantum fluctuations driven by the piston's motion, as mentioned above. Such an effect, of course, dies out at large length scales, as is clear from the figure. This makes the study of quantum heat engines with quantum non-equilibrium particularly interesting at low (mesoscopic) length scales.

\section{Conclusion}
To conclude, we have analytically and numerically determined the efficiency fluctuations in quantum Carnot and Otto engines with quantum pistons operating under non-quasi-static adiabatic expansion and compression processes at constant speed, along with quasi-static isothermal (for the Carnot engine) or isochoric processes (for the Otto engine). We have considered the two-point measurement scheme during the adiabatic strokes in this regard. By considering each of the pistons as a single quantum particle in a 1-D box with a non-quasi-statically movable boundary, we capture the stochastic nature of the work output resulting from the unitary evolution of the quantum mechanical state of the system during non-equilibrium processes. We have explicitly shown how the piston speed during quantum non-equilibrium strokes introduces additional quantum fluctuations that modify the average efficiency of both the engines. We also have analysed the average efficiency for different piston speeds of both the quantum engines. We have obtained the wel-known classical equilibrium results for both the engines in the quasi-static ($v\rightarrow0$, $\tau\rightarrow\infty$, $v\tau=finite$) and the classical (high temperature) limits. 

Eqns. (\ref{eqn16}), (\ref{eqn16a}), (\ref{eqn17}), and (\ref{eqn20}), and figures \ref{fig6}, and \ref{fig4} are the key results obtained by us for both the quantum Carnot and Otto engines. Our findings would be useful for Carnot and Otto engines at mesoscopic scales.

Our results, however, can be generalized by considering many-particle systems (bosons and fermions) as the working substances in the pistons; we leave this as an open problem. A relativistic treatment of this system is likewise left for future work. Furthermore, while we have assumed quasi-static processes for the non-adiabatic strokes, considering non-quasi-static cases would introduce heat input fluctuations, driving the quantum engines far from equilibrium. We leave the study of efficiency fluctuations in such far-from-equilibrium regimes as an open problem.

\section*{Author Contribution}
BKB framed the problem, solved the problem, and wrote the article. SB framed the problem, solved the problem, and wrote the article.

\section*{Data availability statement}
No new data were created or analysed in this study.
\section*{Acknowledgement}
S. Biswas acknowledge the partial financial support of the ANRF, DST, Govt. of India under the EMEQ Scheme [No. EEQ/2023/000788]. Useful discussions with Prof. J. K.  Bhattacharjee (IIT-Kgp, India) and Dr. Saugata Bhattacharyya (VC, Kolkata, India) are gratefully acknowledged.


\begin{thebibliography}{99}
\bibitem{Seifert2012}U. Seifert, \textit{Stochastic thermodynamics, fluctuation theorems and molecular machines}, \href{https://doi.org/10.1088/0034-4885/75/12/126001}{Rep. Prog. Phys. \textbf{75}, 126001 (2012)}
\bibitem{Kosloff2014}R. Kosloff and A. Levy, \textit{Quantum heat engines and refrigerators: continuous devices}, \href{https://doi.org/10.1146/annurev-physchem-040513-103724}{Annu. Rev. Phys. Chem. \textbf{65}, 365 (2014)}
\bibitem{Vinjanampathy2016}S. Vinjanampathy and J. Anders, \textit{Quantum thermodynamics}, \href{https://doi.org/10.1080/00107514.2016.1201896}{Contemp. Phys. \textbf{57}, 545 (2016)}
\bibitem{Binder2018}F. Binder, L. A. Correa, C. Gogolin, J. Anders, and G. Adesso (ed.), \textit{Thermodynamics in the Quantum Regime: Fundamental Aspects and New Directions} (Springer, Cham, 2018)



\bibitem{Talkner2020}P. Talkner and P. Hänggi, \textit{Colloquium: Statistical mechanics and thermodynamics at strong coupling: quantum and classical}, \href{https://doi.org/10.1103/RevModPhys.92.041002}{\rmp{92}{041002}{2020}}; Erratum: \href{https://doi.org/10.1103/RevModPhys.94.039901}{\rmp{94}{039901}{2022}}
\bibitem{Landi2021}G. T. Landi and M. Paternostro, \textit{Irreversible entropy production: from classical to quantum}, \href{https://doi.org/10.1103/RevModPhys.93.035008}{\rmp{93}{035008}{2021}}
\bibitem{Pekola2021}J. P. Pekola and B. Karimi, \textit{Colloquium: Quantum heat transport in condensed matter systems}, \href{https://doi.org/10.1103/RevModPhys.93.041001}{\rmp{93}{041001}{2021}}
\bibitem{Bhattacharjee2021}S. Bhattacharjee and A. Dutta, \textit{Quantum thermal machines and batteries}, \href{https://doi.org/10.1140/epjb/s10051-021-00235-3}{\epjb{94}{239}{2021}}
\bibitem{Cangemi2024}L. M. Cangemi, C. Bhadra, A. Levy, \textit{Quantum engines and refrigerators}, \href{https://doi.org/10.1016/j.physrep.2024.07.001}{Phys. Rep. \textbf{1087}, 1 (2024)}
\bibitem{Campaioli2024}F. Campaioli, S. Gherardini, J. Q. Quach, M. Polini, G. M. Andolina, \textit{Colloquium: Quantum batteries}, \href{https://doi.org/10.1103/RevModPhys.96.031001}{\rmp{96}{031001}{2024}}
\bibitem{Campbell2026}S. Campbell \textit{et al}, \textit{Roadmap on quantum thermodynamics}, \href{https://doi.org/10.1088/2058-9565/ae1e27}{Quantum Sci. Technol. \textbf{11}, 012501 (2026)}



\bibitem{Rossnagel2016}J. Rossnagel, S. T. Dawkins, K. N. Tolazzi, O. Abah, E. Lutz, F. Schmidt-Kaler, and K. Singer, \textit{A single-atom heat engine}, \href{https://doi.org/10.1126/science.aad6320}{\sc{352}{325}{2016}}

\bibitem{Peterson2019}J. P. S. Peterson, T. B. Batalhão, M. Herrera, A. M. Souza, R. S. Sarthour, I. S. Oliveira, and R. M. Serra, \textit{Experimental characterization of a spin quantum heat engine}, \href{https://doi.org/10.1103/PhysRevLett.123.240601}{\prl{123}{240601}{2019}}

\bibitem{Lindenfels2019}D. von Lindenfels, O. Gräb, C. T. Schmiegelow, V. Kaushal, J. Schulz, Mark T. Mitchison,  John Goold, F. Schmidt-Kaler, U. G. Poschinger, \textit{Spin heat engine coupled to a harmonic-oscillator flywheel}, \href{https://doi.org/10.1103/PhysRevLett.123.080602}{\prl{123}{080602}{2019}}

\bibitem{Klatzow2019}J. Klatzow, J. N. Becker, P. M. Ledingham, C. Weinzetl, K. T. Kaczmarek, D. J. Saunders, J. Nunn, I. A. Walmsley, R. Uzdin, \textit{et al}, \textit{Experimental demonstration of quantum effects in the operation of microscopic heat engines}, \href{https://doi.org/10.1103/PhysRevLett.122.110601}{\prl{122}{110601}{2019}}

\bibitem{Lisboa2022}V. F. Lisboa, P. R. Dieguez, J. R. Guimarães, J. F. G. Santos, and R. M. Serra, \textit{Experimental investigation of a quantum heat engine powered by generalized measurements}, \href{https://doi.org/10.1103/PhysRevA.106.022436}{\pra{106}{022436}{2022}}

\bibitem{Koch2023}J. Koch, K. Menon, E. Cuestas, \textit{A quantum engine in the BEC–BCS crossover}, \href{https://doi.org/10.1038/s41586-023-06469-8}{\nat{621}{723}{2023}}

\bibitem{Denzler2024}T. Denzler, J. F. G. Santos, E. Lutz, and R. M. Serra, \textit{Nonequilibrium fluctuations of a quantum heat engine}, \href{https://doi.org/10.1088/2058-9565/ad6287}{Quantum Sci. Technol. \textbf{9}, 045017 (2024)}

\bibitem{Barontini2025}G. Barontini, \textit{Quantum zeno engines and heat pumps}, \href{https://doi.org/10.1103/PhysRevLett.134.010407}{\prl{134}{010407}{2025}}

\bibitem{Hou2025}W. Hou, W. Yao, X. Zhao, K. Rehan, Y. Li, Y. Li, E. Lutz, Y. Lin, and J. Du, \textit{Combining energy efficiency and quantum advantage in cyclic machines}, \href{https://doi.org/10.1038/s41467-025-60179-5}{Nat. Commun. \textbf{16}, 5127 (2025)} 

\bibitem{Uusnakki2026}T. Uusnäkki, T. Mörstedt, W. Teixeira, M. Rasola, and M. Möttönen, \textit{Initial demonstration of a quantum heat engine based on dissipation-engineered superconducting circuits}, \href{https://doi.org/10.1038/s41467-026-72651-x}{Nat. Commun. \textbf{17}, 6054 (2026)}

\bibitem{Xue2026}Q.-F. Xue, Q. Zhang, X.-C. Zhuang, Y.-J. Zhang, Y.-J. Xia, E. Russo, G. Chiribella, R. L. Franco, and Z.-X. Man, \textit{Anomalous heat flows and quantum Otto engine with (in)definite causal order}, \href{https://doi.org/10.1103/sx1m-pdhz}{\prl{137}{030404}{2026}}

\bibitem{Scovil1959}H. E. D. Scovil and E. O. Schulz-DuBois, \textit{Three-level masers as heat engines}, \href{https://doi.org/10.1103/PhysRevLett.2.262}{\prl{2}{262}{1959}}

\bibitem{Alicki1979}R. Alicki, \textit{The quantum open system as a model of the heat engine}, \href{https://doi.org/10.1088/0305-4470/12/5/007}{J. Phys. A: Math. Gen. \textbf{12}, L103 (1979)}

\bibitem{Kieu2004}T. D. Kieu, \textit{The second law, Maxwell's demon, and work derivable from quantum heat engines}, \href{https://doi.org/10.1103/PhysRevLett.93.140403}{\prl{93}{140403}{2004}}.

\bibitem{Quan2007}H. T. Quan, Y. Liu, C. P. Sun, and F. Nori, \textit{Quantum thermodynamic cycles and quantum heat engines}, \href{https://doi.org/10.1103/PhysRevE.76.031105}{\pre{76}{031105}{2007}}

\bibitem{Quan2009}H. T. Quan, \textit{Quantum thermodynamic cycles and quantum heat engines. II.}, \href{https://doi.org/10.1103/PhysRevE.79.041129}{\pre{79}{041129}{2009}}

\bibitem{Kim2011}S. W. Kim, T. Sagawa, S. de Liberato, and M. Ueda, \textit{Quantum Szilard engine}, \href{https://doi.org/10.1103/PhysRevLett.106.070401}{\prl{106}{070401}{2011}}

\bibitem{Horodecki2013}M. Horodecki and J. Oppenheim, \textit{Fundamental limitations for quantum and nanoscale thermodynamics}, \href{https://doi.org/10.1038/ncomms3059}{Nat. Commun. \textbf{4}, 2059 (2013)}

\bibitem{Xiao2015}G. Xiao and J. Gong, \textit{Construction and optimization of a quantum analog of the Carnot cycle}, \href{https://doi.org/10.1103/PhysRevE.92.012118}{\pre{92}{012118}{2015}}

\bibitem{Zheng2016}Y. Zheng, P. Hänggi, and D. Poletti, \textit{Occurrence of discontinuities in the performance of finite-time quantum Otto cycles}, \href{https://doi.org/10.1103/PhysRevE.94.012137}{\pre{94}{012137}{2016}}

\bibitem{Millen2016}J. Millen and A. Xuereb, \textit{Perspective on quantum thermodynamics}, \href{https://doi.org/10.1088/1367-2630/18/1/011002}{\njp{18}{011002}{2016}}




\bibitem{Kosloff2017}R. Kosloff and Y. Rezek, \textit{The quantum harmonic Otto cycle},  \href{https://doi.org/10.3390/e19040136}{Entropy \textbf{19}, 136, 2017}

\bibitem{Watanabe2017}G. Watanabe, B. P. Venkatesh, P. Talkner, and A. del Campo, \textit{Quantum performance of thermal machines over many cycles}, \href{https://doi.org/10.1103/PhysRevLett.118.050601}{\prl{118}{050601}{2017}}

\bibitem{Ding2018}X. Ding, J. Yi, Y. W. Kim, and P. Talkner, \textit{Measurement-driven single temperature engine}, \href{https://doi.org/10.1103/PhysRevE.98.042122}{\pre{98}{042122}{2018}}

\bibitem{Denzler2020}T. Denzler and E. Lutz, \textit{Efficiency fluctuations of a quantum heat engine}, \href{https://doi.org/10.1103/PhysRevResearch.2.032062}{Phys. Rev. Research \textbf{2}, 032062(R), (2020)}

\bibitem{Agarwalla2021}S. Saryal and B. K. Agarwalla, \textit{Bounds on fluctuations for the finite-time quantum Otto cycle}, \href{https://doi.org/10.1103/PhysRevE.103.L060103}{\pre{103}{L060103}{2021}}

\bibitem{Purkayastha2022}A. Purkayastha, G. Guarnieri, S. Campbell, J. Prior, and J. Goold, \textit{Periodically refreshed quantum thermal machines}, \href{https://doi.org/10.22331/q-2022-09-08-801}{Quantum \textbf{6}, 801 (2022)}

\bibitem{Monnai2025}T. Monnai, \textit{General lower and upper bounds of efficiency of heat engines}, \href{https://doi.org/10.1103/PhysRevE.111.044135}{\pre{111}{044135}{2025}}

\bibitem{Svidzinsky2026}A. A. Svidzinsky and W. Zhang, \textit{Quantum evolution of mixed states and performance of quantum heat engines}, \href{https://doi.org/10.1103/v5bc-p2mw}{\pre{113}{054102}{2026}}

\bibitem{Haywood2018}J. B. Heywood, \textit{Internal combustion engine fundamentals}, 2nd ed. (McGraw-Hill Education, New York, 2018)

\bibitem{Quan2012}H. T. Quan and C. Jarzynski, \textit{Validity of nonequilibrium work relations for the rapidly expanding quantum piston}, \href{https://doi.org/10.1103/PhysRevE.85.031102}{\pre{85}{031102}{2012}}

\bibitem{Doescher1969} S. W. Doescher and M. H. Rice, \textit{Infinite square-well potential with a moving wall}, \href{https://doi.org/10.1119/1.1975291}{Am. J. Phys. 37, 1246–1249 (1969)}

\bibitem{Zemansky1997}M. W. Zemansky and R. H. Dittman, \textit{Heat and thermodynamics}, (Sec. 7.1, 7th ed., McGraw-Hill, 1997)


\bibitem{Talkner2007} P. Talkner, E. Lutz, and P. Hänggi, \textit{Fluctuation theorems: Work is not an observable}, \href{https://doi.org/10.1103/PhysRevE.75.050102 }{\pre{75}{050102}{2007}}

\bibitem{Campisi2011}M. Campisi, P. Hänggi, and P. Talkner
\textit{Colloquium: quantum fluctuation relations: foundations and applications}, \href{https://doi.org/10.1103/RevModPhys.83.771}{\rmp{83}{1653}{2011}}

\bibitem{Deffner2008} S. Deffner and E. Lutz, \textit{Nonequilibrium work distribution of a quantum harmonic oscillator}, \href{https://doi.org/10.1103/PhysRevE.77.021128}{\pre{77}{021128}{2008}}

\bibitem{Udo2012} U. Seifert, \textit{Stochastic thermodynamics, fluctuation theorems, and molecular machines}, \href{https://doi.org/10.1088/0034-4885/75/12/126001}{Rep. Prog. Phys. \textbf{75}, 126001 (2012)}; \textit{Stochastic thermodynamics: principles and perspectives}, \href{https://doi.org/10.1140/epjb/e2008-00001-9}{\epjb{64}{423}{2008}}

\bibitem{Biswas2010}S. Biswas, J. K. Bhattacharjee, D. Majumder, K. Saha, and N. Chakravarty, \textit{Casimir force on an interacting Bose–Einstein condensate}, \href{https://doi.org/10.1088/0953-4075/43/8/085305}{\jpb{43}{085305}{2010}}

\bibitem{Dey2020}S. Dey, P. Manchala, S. Basu, D. Banerjee and S. Biswas,\textit{Finite-size effects on the cluster expansions for quantum gases in restricted geometries}, \href{https://doi.org/10.1088/1402-4896/ab8e01}{Phys. Scr. \textbf{95}, 075003, (2020)}
\end{thebibliography}
\end{document}